\documentstyle[11pt,aaspp4]{article}
\newcommand\beq{\begin{equation}}
\newcommand\eeq{\end{equation}}

\begin{document}

\title{Time-Dependent Photoionization in a Dusty Medium II: 
Evolution of Dust Distributions and Optical Opacities}

\author{Rosalba Perna\altaffilmark{1,2}, Davide Lazzati\altaffilmark{3} and Fabrizio Fiore\altaffilmark{4}}

\altaffiltext{1}{Harvard Society of Fellows, 74 Mount Auburn Street, Cambridge, MA 02138}

\altaffiltext{2}{Harvard-Smithsonian Center for Astrophysics, 60 Garden Street,
Cambridge, MA 02138}

\altaffiltext{3}{Institute of Astronomy, University of Cambridge, 
Madingley Road, Cambridge CB3 0HA, UK}

\altaffiltext{4}{Osservatorio Astronomico di Roma, via dell'Osservatorio, 2,
I-00040, Monte Porzio Catone, Roma, Italy}
\begin{abstract}

The interaction of a radiation field with a dusty medium is a relevant
issue in several astrophysical contexts.  We use the time-dependent
photoionization code in a dusty medium developed by Perna \& Lazzati
(2002), to study the modifications in the dust distribution and the
relative optical opacities when a strong X-ray UV radiation flux
propagates into a medium.  We find that silicates are preferentially
destroyed with respect to graphite, and the extinction curve becomes
significantly flatter (hence implying less reddening), with the
characteristic bump at $\lambda$ 2175 $\AA$ highly suppressed, due to
the destruction of the small graphite grains. This could explain the
observational lack of such a feature in GRB afterglow and AGN
spectra. For a very intense and highly variable source irradiating a
compact and dense region, time variability in the optical opacity
resulting from dust destruction can be observed on a relatively short
timescale. We show that, under these circumstances, monitoring the
time variability of the opacity can yield powerful clues on the
properties of dust in the environment of the source. In particular, it
allows to break the observational degeneracy as to whether a grey
extinction is the result of a low dust-to-gas ratio or of a dust grain
distribution that is skewed towards large grains.

\end{abstract}


\section{Introduction}

The effects of the interaction between a strong radiation field and a
dusty environment are of relevance in several astrophysical contexts,
most notably Active Galactic Nuclei (AGNs), Gamma-Ray Bursts (GRBs)
and star formation (SF).  The radiation field can significantly alter
the properties of the dust in the close environment of the source,
producing effects that, for a time-variable and intense source, can be
observable on a very short time scale. When time variability of the
opacities is indeed observed, a great deal of information can be
gained on the properties of the dusty cloud which is being
irradiated.  This is for example the case of GRBs, where time
variability in absorption lines (Perna \& Loeb 1998; B{\"o}ttcher et
al. 1999; Mirabal et al. 2002), in X-ray opacity (Lazzati et al. 2001a,
Lazzati \& Perna 2002), and in optical extinction (Waxman
\& Draine 2000; Draine \& Hao 2002; Perna \& Lazzati 2002), have been
proposed and used (Amati et al. 2000; Lazzati et al. 2001b) to learn
about the properties of the close GRB environment. And it should be
noted that the nature of the GRB environments is a particularly
important issue not only because the type of environment in which GRBs
occur can be a strong diagnostic of the GRB progenitors, but more
generally because GRBs probe inner regions of galaxies, not accessible
to QSO absorption studies. Therefore they constitute an invaluable
complement to QSO absorption studies in order to have a fair idea of
what the properties of high-redshift galaxies are. To this purpose, it
is particularly important to have a proper estimate of metallicity and
dust content (i.e. dust-to-gas ratio).

The properties of dust in the close environment of AGNs appear
somewhat anomalous and, in general, different than the Galactic
ones. Laor \& Draine (1993) note how spectral observations suggest
that silicates are highly depleted with respect to graphite, and/or
dust must be composed of grains larger than the typical grains in the
Galaxy. Similar conclusions on the presence of a dust distribution
which is skewed towards large grains has also been reached by Maiolino
et al. (2001a; see however Weingartner \& Murray 2002), whose analysis
includes measurements of reddening for a large sample of AGNs and
comparison with the Galactic value.

In the case of GRBs, measurements of dust properties have yielded so
far results that may appear contradictory: Savaglio, Fall \& Fiore
(2002) find a dust-to-gas ratio in GRB spectra that is higher than in
QSO-selected DLAs and generally consistent with the solar value, while
Wijers \& Galama (2000), based on an analysis of the reddening in the
afterglow spectra (and using a Galactic-type extinction curve) infer a
dust-to-gas ratio that is much lower than the Galactic value. A way to
reconcile the two types of observations can be the presence of a dust
grain distribution that is skewed towards larger grains (e.g. Stratta
et al. 2002), similarly to what inferred for AGNs.  As far as
extinction is concerned, a Galactic-type grain distribution with a low
dust-to-gas ratio can yield the same extinction of a grain distribution
that is skewed towards larger grains, but for which the dust-to-gas
ratio is larger.

In this paper we discuss the shape and composition of the dust
distribution left behind by a strong and variable source of ionizing
photons, as well as the evolution of the extinction during
the destruction phase. We will show that, since small grains can be
more easily destroyed by the interaction with the photons, and since
graphite particles are more resistant to destruction, the typical
outcome is a dust distribution skewed towards big graphite
particles. Such a result may explain the observational findings of
Laor \& Draine (1993) and Maiolino et al. (2001a). In order to test
such speculations, we suggest to use, as sources, the very early
afterglows of GRBs. In these events, in fact, a virtually unperturbed
cloud is suddenly illuminated by a strong photon flux, and the
evolution of reddening and absorption may be followed through
photometric observations of the dedicated {\em Swift} satellite.  
We will show how monitoring the time evolution of the extinction 
allows to break the observational
degeneracy as to whether a grey extinction is the result of a low
dust-to-gas ratio or of a dust grain distribution that is skewed
towards large grains.
It should
however be remarked that a significant (and therefore observable) time
variability can only be observed for a source which irradiates a dense
and compact region (e.g.  Perna \& Loeb 1998; Mirabal et
al. 2002). For the case of GRBs, this should be possible in some
cases, as there is evidence that at least a fraction of long GRBs are
associated with the collapse of massive stars (Fruchter et al. 1999,
Kulkarni et al. 1999), and therefore are expected to occur in dense
and dusty environments as typical of star forming regions.

The paper is organized as follows: in \S 2, we discuss the theoretical
modelling of the extinction curve and its observational
characteristics as a function of the underlying dust grain
distribution. In \S 3, we study the time evolution of the opacity
under an intense radiation field in different types of dusty
environments, including the influence on the opacity of the presence
of Hydrogen in molecular form. The observational perspectives for
detection of variability on a short time scale are discussed in \S
4. Finally, a summary is presented in \S 5.

\section{Modelling the Evolution of the Dust Grain Distribution}

\subsection{Brief review of the code}

The code which models the evolution of the dust grain distribution is
described in detail in Perna \& Lazzati (2002; Paper I in the
following); here we only provide a brief summary.  The code includes
the destruction processes of dust due to UV and X-ray sublimation, as
well as ion field emission\footnote{For heavily charged grains, the
process of ion field emission (IFE) is in competition with the process
of Coulomb explosion (CE) that results in grain fragmentation (Waxman
\& Draine 2000; Fruchter, Krolik \& Rhoads 2001; Reichart 2001a). The
importance of CE versus IFE is uncertain, as it depends on the not
well known resistance of chemical bonds to disruption by a
photoionization field (see Draine \& Hao 2002 and Paper I for a more
extended discussion). Here, as in Paper I, we assume that IFE is the
dominant effect in highly charged grains. However, we envisage an
observational test (see \S 4) that can help discriminate whether IFE
or CE actually dominate.}.  It treats dust as a mixture of silicates
and graphite (similarly to what inferred for Galactic-type dust;
Mathis et al. 1977), and it follows the evolution of each of the two
species. It allows for any initial grain-size distribution, and it
follows its evolution in space and time. While following the evolution
of the dust grains under the influence of the radiation field, the
code also follows the evolution of Hydrogen (which can initially be in
any mixture of its atomic and molecular phase), and the 12 most
abundant astrophysical elements, using the photoionization routines
described in Perna, Raymond \& Loeb (2000).  For the elements depleted
into dust (i.e. C, Mg, Si, O, Fe), the abundances in the gaseous and
in the dust-depleted phase are separately being kept track of,
following the recycling from dust into gas as dust is being sublimated
away.

\subsection{Computation of the extinction curve}

The dust extinction curve $A(\nu)$ in the region below 1.25 keV is
computed in the code by interpolation using the tables\footnote{They
can be found at {\tt
http://astro.Princeton.EDU/$^\sim$draine/dust/dust.html}.}  provided
by Draine and collaborators (Draine \& Lee 1985; Laor \& Draine 1993)
for the coefficient of absorption $Q_{\rm abs}(a,\nu)$, the
coefficient of scattering $Q_{\rm sca}(a,\nu)$, and the scattering
asymmetry factor $g$.  In terms of these quantities, the extinction in
a region of radius $R$ is given by
\beq 
A(\nu)=\int_0^R dr\,\sum_{i=1}^2\int da\;\pi a^2 \frac{dn_i(a)}{da}[Q_{\rm
abs,i}(a,\nu)+(1-g)Q_{\rm sca,i}(a,\nu)]\;.
\label{eq:Anu}
\eeq
In the above equation, $dn_i/da$ indicates the dust grain
distribution, while the indices $i=1,2$ represent the two types of
dust grains that we are considering here, i.e. silicate and graphite.
For simplicity we assume that, prior to the onset of the source,  
the size distribution of the grains is
a power law\footnote{This is a good approximation for our Galaxy,
while not much is known on the shape of the distribution in other
galaxies.}
\beq
\frac{dn_i}{da}=A_i n_{\rm H}\,a^{-\beta}\;\;\;\;\;\;\; a_{\rm min}\le
a\le a_{\rm max}\;.
\label{eq:dnda}
\eeq 
For a given size distribution of the dust grains (characterized by the
parameters $a_{\rm min}$, $a_{\rm max}$ and $\beta$), the coefficients
$A_i$ can be related to the dust-to-gas ratio $f_d$ (defined as the
ratio between the total mass in dust and the total mass in hydrogen)
through the relation
\beq
f_d\equiv\frac{m_{\rm dust}}{m_{\rm Hyd}} = \frac{4\pi}{3 m_{\rm H}}
\frac{a_{\rm max}^{4-\beta}}{(4-\beta)}
\left[1-\left(\frac{a_{\rm min}}{a_{\rm max}}\right)^{4-\beta} \right]
\sum_i A_i\rho_i\;,
\label{eq:fd}
\eeq
having assumed a spherical shape for the dust grains.  The value
$f_d\approx 0.01$ is typical of the solar neighborhood with the grain
densities $\rho_{\rm Sil}\approx 3.3$ g cm$^{-3}$ and $\rho_{\rm
Gra}\approx 2.26$ g cm$^{-3}$ (e.g. Draine \& Lee 1984), respectively
for silicates and graphite. We take the initial mass ratio between the
two species of grains to be similar to the solar value
\begin{equation}
m_{\rm G_S}={{A_{\rm Gra}\,\rho_{\rm Gra}}
\over{A_{\rm Sil}\,\rho_{\rm Sil}}} \approx 0.88\;.
\label{eq:mdr}
\end{equation}

Figure 1 shows the extinction curve constructed as described in
Eq.~(\ref{eq:Anu}), and using the grain distribution in
Eq.~(\ref{eq:dnda}) for the size range $a_{\rm min}= 0.005 \,\mu {\rm
m}$, $a_{\rm max}= 0.25\, \mu {\rm m}$, which best describes the
Galactic absorption with a slope $\beta=3.5$ (Mathis et al. 1977).
The Galactic extinction curve (solid line) is compared in Figure 1 to
extinction curves produced by shallower distributions of grains. As it
can be seen, the shallower the grain size distribution, the flatter
the extinction curve (and therefore the smaller the reddening).  This
is shown more clearly in Figure 2, where the extinction in the V band
(top panel) and the B-V reddening (bottom panel) are shown as a
function of $\beta$ for three different values of the maximum size of
the grain distribution $a_{\rm max}$.

\subsection{Time evolution of the grain size distribution and extinction curve}

As summarized in \S 2.1 (and described in great detail in Paper I),
our code follows in space and in time the evolution of the grain
distribution $dn_i/da$.  An example is shown in Figure 3, where the
shape of the grain distribution (for the silicates and the graphite
separately) at some distance $r\sim 2\times 10^{18}$ cm from the
source is plotted at several times while the region is being
illuminated by a source with spectrum $L_\nu\propto\nu^{-0.5}$ and
normalization equal to $10^{50}$ erg/sec in the 1eV - 100keV
range\footnote{It should be remarked that, unlike photoionization, for
the purpose of dust destruction what matters is not only the fluence
of the source, but also its luminosity. Therefore, for the same
fluence, a source of shorter duration but higher luminosity is more
effective to destroy dust than a less luminous but more longer-lived
source. This is why the prompt emission of GRBs is so effective in
destroying dust.}.  At $t=0$, the (power-law) grain distribution has a
slope $\beta=3.5$ in the size range \{0.005, 0.25\} $\mu$m.  As grain
destruction proceeds, the value of the largest grain in the
distribution, $a_{\rm max}$, is being reduced. The grain distribution
evolves as grains are being reduced in size, thus causing an increase
in the density of particles of smaller size $a$. On the other hand,
the smaller the grains, the more easily they are sublimated away, and
therefore the low-$a$ tail of the distribution evolves more rapidly.
The overall effect of the time-evolution is a gradual flattening of
the distribution and reduction of $a_{\rm max}$.  It should also be
noted that the grain distribution for the silicates evolves much more
rapidly than that for graphite, which is harder to destroy (see also
Paper I).

Figure 4 (top panel) shows the evolution of the extinction curve
$A(\nu)$ for the same illuminating spectrum used in Figure 3, and for
a region of size $R=10^{19}$ cm and density $n=10^3$ cm$^{-3}$, with a
dust-to-gas ratio equal to the solar value.  The initial size
distribution, $dn_i(r,t=0)/da$, of the grains producing the extinction
is also taken as in Figure 3, for all $r\le R$.  As grain destruction
proceeds, the extinction curve becomes flatter, reflecting the
shallower slope of the grain size distribution, as illustrated by
Figure 3. Another important consequence of the modification in the
grain size distribution is a significant reduction of the
characteristic bump at $\lambda$ 2175 $\AA$ in the extinction curve. This
bump is indeed found to be highly depleted in AGNs (see e.g. Maiolino
et al. 2001b), and it has not been found in spectroscopic analysis of
GRB light curves, even though the resolution should have been
sufficient to detect it (Galama \& Wijers 2001).

The bottom panel of Figure 4 shows the relative contribution to the
opacity from graphite and silicates.  Before the onset of the source,
the dust mass in these two components is taken as for galactic-type
dust (Mathis et al. 1977).  However, it can be seen that, as the radiation
flux impinges on the dust grains, the contribution to the opacity by
silicates becomes negligible with respect to that by graphite.  This
is due to the shorter timescale for destruction of silicates (see
Figure 3 and Paper I). 

An important implication of this result in the context of GRBs
occurring in dense and compact regions (only for these regions, we
remark once again, dust can be significantly destroyed by the burst
prompt emission) is that afterglow observations, which are made at
later times, would preferentially detect the absorption features in
the spectra associated with graphite with respect to those associated
with silicates.  However, as the absorption features are mostly
produced by small grains, and these are preferentially destroyed by
the radiation also in the case of graphite, there will be an overall
suppression of the absorption features associated with graphite as
well (and this is exemplified by the suppression of the $\lambda$ 2175
dust bump discussed above).  More generally, our results imply that,
in a dusty medium subject to an intense source of radiation, the
properties of dust must be different than the Galactic one, in that
the silicates should be depleted with respect to graphite. 
It should be noted that, for
the spectrum assumed in this discussion, dust destruction is mainly
due to thermal sublimation of the dust grains, heated by absorption of
the UV continuum. Nevertheless, this result seems to be quite
independent of the spectrum of the ionizing source, and holds even for
an IFE dominated destruction (see Lazzati \& Perna 2002, Paper 3).

We speculate that our results could help explain the unusual depletion
of silicates with respect to graphite also in the environment of
active galactic nuclei (e.g. Laor \& Draine 1993). In the Maiolino et
al. (2001a, 2001b) sample, the central source of photons has
[2--10]~keV X-ray luminosities ranging from $10^{40}$ to
$10^{45}$~erg~s$^{-1}$. If such sources undertake fluctuations in the
luminosity of one order of magnitude or more on a timescale shorter
than the dust creation timescale\footnote{The time scale for metals 
to return from the gas to grow large grains has been estimated by
Laor \& Draine (1993) to be $\sim 3\times 10^4 n_6^{-1} \Delta v_5^{-1}$ yr,
where $n_6\equiv n/10^6$ cm$^{-3}$ is the density of the region, and
$\Delta v_5^{-1}\la 50$ is the mean speed of metal ions relative to the grains.}, 
a grain distribution skewed towards large grains and dominated
by graphite particles would be produced at a distance $R\sim10^{17}
L_{46}$~cm from the ionizing source, where $L_{46}$ is the
[1~eV--100~keV] peak luminosity in units of $10^{46}$~erg~s$^{-1}$. It
is worth mentioning, however, that the grains with size as large as
1~$\mu$m discussed by Maiolino et al. (2001b) could not be created,
unless already present before the sublimation took place.

\section{Time evolution of opacities in various environments}

In this section we explore the time evolution of the opacities under
an intense X-ray UV radiation field in various types of environments,
characterized by different types of dust distributions and various
fractions of Hydrogen in its molecular form, H$_2$.  We show how the
time evolution can be a diagnostics of these quantities. We adopt here
(unless otherwise noted) an illuminating spectrum
$L_\nu\propto\nu^{-\alpha}$, with $\alpha=0.5$ and normalization as in
\S 2.3.  The dependence of the opacities on the spectrum have been
discussed in Paper I.

\subsection{Time-variable extinction and reddening for various 
grain distributions}

As discussed in the Introduction, measurements of spectral reddening
at one given time can yield similar results for Galactic-type grain
distributions with a low dust-to-gas ratio $f_d$, and for grain
distributions skewed towards larger grains but with an higher value of
$f_d$. Here we show that this degeneracy in the interpretation of the
observational results can be removed with early time/continuous
monitoring of the opacity and reddening.

The behavior of the optical opacity in the V band, $A_{\rm V}$, is
shown in the top panel of Figure~5 for three different values of the
dust-to-gas ratio, that is $f_d=0.01\;f_\odot,\;0.1f_\odot,$ and
$f_\odot$.  For the value $f_d=0.01f_\odot$, we adopted in the
simulation a slope $\beta=3.5$ and a maximum grain size $a_{\rm max}=0.25\mu{\rm m}$,
as typical of our Galaxy.  For the
other two values of $f_d$ we chose, for the corresponding simulation (and using the
same value of $a_{\rm max}$),
the slope $\beta$ that yielded the same reddening of the model
\{$f_d=0.01f_\odot$, $\beta=3.5$\}.  In all cases, the region has
density $n=10^3$ cm$^{-3}$ and radius $R=10^{19}$ cm.  For an observer
that were to infer the values of $f_d$ and $\beta$ based on reddening,
the three physical situations would be degenerate and not
distinguishable on an observational basis. However, as Figure 5 shows,
the time evolution of the opacity under the influence of an intense
radiation field is different in the various scenarios. In particular,
the opacity evolves more rapidly for steeper distributions. This is
due to the fact that smaller grains are sublimated more easily (see
e.g. Paper I).

The time evolution of the reddening, $E_{\rm B-V}$, for the same
values of \{$f_d$, $\beta$\} is shown in the bottom panel of Figure
5. What is interesting to note here is the fact that, while $A_{\rm
V}$ evolves more rapidly for {\em steeper} grain distributions,
$E_{\rm B-V}$ evolves more rapidly for {\em shallower} grain
distributions.  To understand why this is so, one needs to look at
Figure 2 which shows the behavior of $A_{\rm V}(\beta)$ and E$_{\rm
B-V}(\beta)$. While $A_{\rm V}(\beta)$ is almost insensitive to
$\beta$ (for the $a_{\rm max}=0.25\mu{\rm m}$ case that we have assumed here), 
E$_{\rm B-V}(\beta)$ is a very sensitive function of $\beta$,
and in particular it varies more and more rapidly as $\beta$
decreases. This has the implication that, while the time-variable
behaviour of $A_{\rm V}(t)$ is essentially determined by the fact that
the distribution with larger $\beta$ evolves more rapidly (because of
the more efficient destruction of the smaller grains), the behaviour
of E$_{\rm B-V}(t)$ is dominated by the fact that E$_{\rm B-V}$
decreases much more rapidly with $\beta$ for smaller $\beta$.

The trend in the time evolution of the V-band extinction and reddening
for different slopes of the initial grain distribution discussed above
(for a region of density $n=10^3$ cm$^{-3}$ and radius $R=10^{19}$~cm),
still remains for a larger but less dense region (characterized
by the same column density $N_{\rm H}=10^{22}$ cm$^{-2}$). This is
shown in Figure 6, where the region is characterized by a lower
density $n=10^2$ cm$^{-3}$ but larger radius $R=10^{20}$ cm. The
overall rate of time variation is however smaller due to the reduction
in the flux at the larger radii.

It should be noted that, for a given spectrum\footnote{The degree of
time variability due to dust destruction also depends on the spectral
index of the illuminating radiation (see e.g. Paper I); however, the
spectrum is a quantity that is measured independently and therefore it
does not constitute an element of degeneracy for this type of
observations. }, there is a certain degree of degeneracy (in the
observed time-variability in a given band) between the size/density of
the region which is being illuminated and the spectral slope of the
dust-grain distribution in the region.  However, this degeneracy can
be broken by measurements of time variability in both the optical and
the X-ray band. In fact, while the former is mostly (but not only)
governed by the process of dust destruction, the latter is mostly
(but, again, not only; see Paper III), governed by the process of
photoionization of the metals.  And it should also be reminded that,
as a result of the simultanous processes of dust destruction and metal
photoionization, the dust-to-gas ratio that is inferred from a
comparison between the optical and X-ray extinction is most likely
not reflecting the intrinsic properties of the region (see Paper I for
a detailed discussion of this issue).

Figure 7 shows the time dependence of the V-band extinction and
B-V reddening for grain size distributions with the same slope $\beta=3.5$
but different values of the maximum size $a_{\rm max}$ of the grains
in the distribution. In all cases, the region has a density $n=10^3$
cm$^{-3}$ and radius $R=10^{19}$~cm with a solar value for the
dust-to-gas ratio. Both the V-band extinction and the reddening evolve
more slowly for grain distributions with larger $a_{\rm max}$, due to
the longer sublimation time of the larger grains (see Paper I).

It should be remarked that the results above have been given for a
source of constant luminosity in the time interval \{0, $t$\}.  In the
case where the source is the prompt GRB emission then, if no
monitoring of the optical opacities is made during the prompt phase,
the extinction and reddening that an observer would measure in the
later afterglow phase is what it is at the time at which the prompt
X-ray UV flash ends. This has the implication that short bursts are
less likely to affect their environments than long bursts.  Therefore,
if the long GRBs with detected optical counterparts are actually those
which manage to evacuate a funnel in their dusty environment, as
claimed by Reichart (2001b), then it will be more difficult to detect
optical counterparts for short bursts.  Another consideration is the
fact that, if short bursts are associated with mergers of two compact
objects, and occur in low-density environments (e.g. Bloom, Sigurdsson
\& Pols 1999; but see Perna \& Belczynski 2002), then no variability
should be observable for them with early-time monitoring of the
opacities. These types of observations could therefore provide an
independent diagnostics of whether the classes of long and short
bursts are actually associated with two different types of
progenitors.

\subsection{Effects of the presence of H$_2$ on the time-variable extinction}

The presence of molecular Hydrogen in a cloud can be revealed through
the absorption signatures imprinted in the spectrum (due to
vibrationally excited H$_2$) in the wavelength range $1110 \AA<\lambda < 1705 \AA$.  A detailed
computation of the trasmission spectrum has been performed by Draine
\& Hao (2002). What we are mainly interested in here is the effect of
the presence of molecular Hydrogen on the time variable signatures
discussed above produced by various dust distributions.

Hydrogen in molecular phase can alter the evolution of the dust
destruction fronts essentially for two reasons. Firstly, the cross
section to photoionize H$_2$ and H$_2^+$ is different than that of
Hydrogen; therefore, unless the absorbing region is optically thin to
H and H$_2$, the propagating flux (which is also responsible for dust
sublimation) will be modified differently depending on whether
Hydrogen is in its atomic or molecular form. Second and most
important, is the presence of the singly ionized molecular Hydrogen
H$_2^+$. It is generated by photoionization of H$_2$, and it is
destroyed by the processes of photoionization (H$_2^+ +\, h\nu
\rightarrow$ 2H$^+ + e^-$) and photodissociation (H$_2^+ + h\nu
\rightarrow$ H$^+ +\, $H).  While both the cross section thresholds to
photoionization of  H$_2$ and H$_2^+$ are above that of Hydrogen,
the cross section for the process of photodissociation of H$_2^+$ is given by
\beq
\sigma(E)=2.7\times 10^{-16} {\rm cm}^2 \left(\frac{E}{29\,{\rm eV}}\right)^2
\left(1-\frac{E}{29\, {\rm eV}}\right)^6\;\;\;\;\;\; {\rm for} E< 29\; {\rm eV}\;,
\label{eq:h2p}
\eeq
where we have adopted the fit derived by Draine \& Hao (2002) to the
photodissociation cross section found by von Bush \& Dunn (1972) after
averaging over the H$_2^+$ vibrational distribution.  It can be seen
that the cross section in Eq.~(\ref{eq:h2p}) gives a direct
contribution to the opacity in the optical range.

Let us now assume that, at $t=0$, all Hydrogen is in its molecular
form with an initial density $n({\rm H}_2)=n_{\rm H}/2$.  As discussed
above, the process of complete photoionization of H$_2$ passes though
the intermediate state of H$_2^+$. During the process of
photoionization, this is formed in a rather thin layer behind the
photoionization front of H$_2$.  Draine \& Hao (2002) found in their
simulations that the column density $N$(H$_2^+$) quickly stabilizes at
a constant value that is roughly independent of the density of the
medium, while it depends on the hardness of the incident spectrum,
being lower for softer spectra.  Our simulations confirm their
results. Figure 8 shows the contribution to the V band opacity (top
panel) and to the B-V reddening (bottom panel) due to the H$_2^+$
column density. We show the results for various combinations of values
of the medium density ($n_{\rm H}=10$ cm$^{-3}$, $n_{\rm H}=10^{2}$
cm$^{-3}$ and $n_{\rm H}=10^{3}$ cm$^{-3}$) and of the spectrum
hardness ($\alpha=0.5$ and $\alpha=1$). As it can be seen from the
Figure (note that the time oscillations are due to numerical
inaccuracies in traversing the radial zones), H$_2^+$ provides a
roughly constant value to the opacities, that is approximately
independent of the density of the region and that is lower for a
softer spectrum.

If molecular Hydrogen is present in the absorbing region, then it will
be possible to appreciate the time variability in the opacity due to
dust destruction only if the dust opacity is larger (at $t=0$ and for
at least a fraction of the time during which dust destruction takes
place) than the contribution to the opacity provided by
H$_2^+$. Figure 9 shows the regions in the parameter space \{$\beta,f_d$\}
of values of $\beta,f_d$ such that the contribution to opacity and
B-V reddening by dust at $t=0$ is above the contribution due to H$_2^+$.
The shaded region in the Figure denotes the parameter space where both
$A_V(t=0)$ and $E_{\rm B-V}(t=0)$ due to dust are above the
corresponding contributions due to H$_2^+$.

If the H$_2^+$ contribution to the opacity is lower than that of dust,
then the time variable signatures due to dust destruction can be more
easily identified. The differences in the trasmitted spectrum
impinging on the dust grains, due to differences in the opacities of H
and H$_2$, affect only marginally the time evolution of the dust
opacity.  This is illustrated in Figure 10 which shows, for two types
of initial dust distributions, a comparison between the time
variability of the dust contribution to the V band extinction and
reddening, when all Hydrogen is in its atomic phase and when all
Hydrogen is in its molecular phase before the onset of the source.
 
\section{Observational prospects for observing time-variability during GRBs}

Time-variable extinction can be detected when dust is being destroyed
on a timescale comparable to the observation time. In order for this
to happen, a source with very high X-ray/UV/optical luminosity is
needed. The radiation generated by the prompt GRB emission constitutes
the best example known so far of such a high luminosity source.

Optical observations of GRB afterglows are now performed regularly,
with a reaction time that is decreasing with time.  The time
variability that we have described, however, requires optical
monitoring {\em during} the early prompt phase. This constraint cannot
be overemphasized. In fact, even if the afterglow phase of a GRB can
have a fluence comparable to the prompt emission in the UV and soft
X-ray range, dust sublimation will take place almost only during the
prompt phase. This is due to the fact that, since the temperature of a
grain is due to the balance of an input term (the rate of UV photon
absorption) and an output one (the thermal emission of the grain),
what causes the sublimation is the luminous prompt phase rather than
the long lasting afterglow phase. Such a constraint is not applicable
for photoionization, which proceeds as a function of the fluence and
not of the flux, and therefore is much easier to detect the
evaporation of the soft X-ray column than that of dust in the
afterglow of GRBs (see Paper 3 for a more thorough discussion).  
The IFE dust destruction process, on the other hand, is sensitive to
the fluence and may be relevant also in the afterglow phase. Numerical
simulations in IFE dominated situations (see Paper 3), suggest however
that IFE can only marginally contribute to the dust destruction even
for hard GRB spectra. This consideration allows us to suggest a test
to asses the importance of CE fragmentation, which is not included in
the code (being exclusive with IFE). If CEs take place, the X-ray flux
of the afterglow has still enough ionizing photons to modify the dust
distribution and column density. For this reason, should any evolution
of the extinction or reddening be observed during the afterglow phase
of GRBs, this should be ascribed to the effect of Coulomb explosions, which
are more effective than IFE and sensitive to fluence. To date, such an
effect has never been observed, even though better data are certainly
to be awaited before firm conclusions can be drawn.

Monitoring of the burst optical-UV flux during the prompt phase will
be possible with {\em Swift}, which is planned to be launched in
2003. The optical telescope on board {\em Swift} will in fact provide
optical and UV photometry and spectra on a timescale on the order of a
few tens of seconds after the GRB trigger. In a few cases, the
repointing of the optical telescope should be made in even less time.
For a GRB at $z\sim 1$, the V band rest frame falls in the NIR.
Therefore, for the studies that we have proposed here, particularly
important are the planned robotic telescopes from the ground (e.g. the
{\rm REM} telescope, Zerbi et al. 2002), which will provide NIR
photometry on the same short timescale as {\em Swift} for the optical
band.

\section{Summary}

We have used the time-dependent photoionization code for a dusty
medium, developed by Perna \& Lazzati (2002), to study the
observational signatures in the optical extinction due to the
evolution of the dust distribution in the close environment of a
source of intense radiation field. This is of relevance in several
astrophysical contexts, such as AGNs, GRBs and, albeit at different
scales, for star formation. We have shown that a strong X-ray
UV/optical radiation field can significantly modify the dust
properties in the close environment of the source. In particular, we
have found that the dust grain distribution becomes significantly
flatter as a result of the faster sublimation of the smaller
grains. This has the important consequence that the characteristic
dust ``bump'' at 2275 $\AA$ is highly depleted. The lack of such a feature
has been noted both in observations of AGNs and GRBs. Another
consequence of the interaction of the radiation field with the dusty
medium is the preferential destruction of silicates with respect to
graphite. This might help explain the observed depletion of silicates
with respect to graphite in AGNs.

In the case of a very luminous source, time variability can be
detected on a relatively short timescale.  This is the case of GRB
sources occurring in dense and compact regions.  We have shown how
early-time (e.g. during the prompt phase) monitoring of the optical
opacities with {\em Swift} and {\em REM} can shed light on the type of
dust (in particular with respect to its grain-size distribution) in
the environment of GRBs. More specifically, we have shown how it
allows to break the observational degeneracy as to wheather a grey
extinction is the result of a low dust-to-gas ratio or of a dust grain
distribution that is skewed towards large grains.  The observations
that we have proposed here are particularly relevant in order to gain
a fair knowledge of what the properties (such as metallicity and dust
content) of high-redshift galaxies really are.

\acknowledgements We thank Roberto Maiolino for insightful comments on our
manuscript. RP thanks the Osservatorio Astronomico di Monte Porzio-Roma
for its kind hospitality and financial support during the time that part
of this work was carried out.

\begin{figure}
\centerline{\epsfysize=5.7in\epsffile{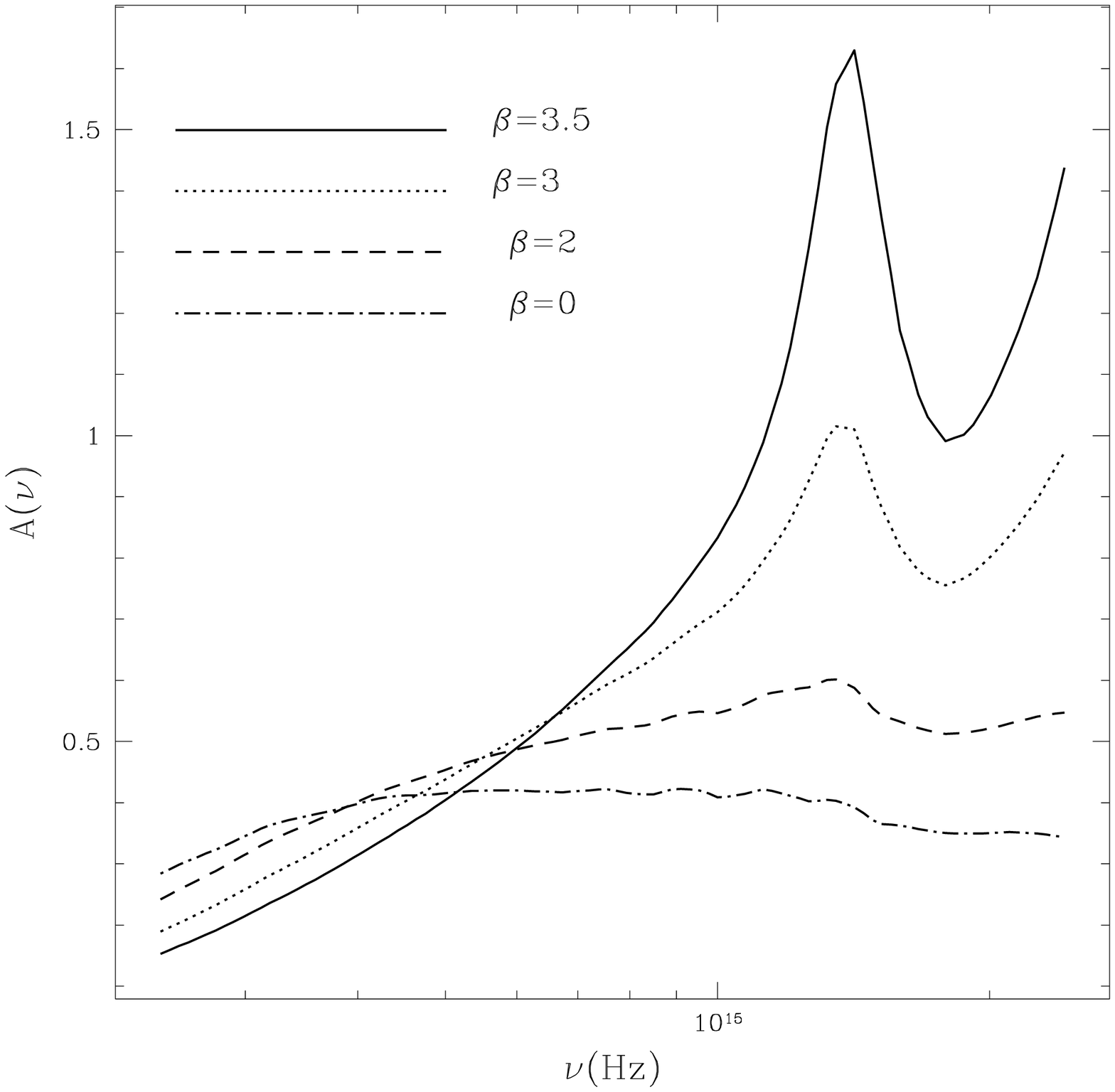}}
\caption{The extinction curve in the 1-10 eV energy range for
power-law grain distributions of various slopes $\beta$ and 
grains in the range \{$a_{\rm min}=0.005 \mu {\rm m}$, $a_{\rm max}=0.25
\mu {\rm m}$\}. The Hydrogen column density is $N_{\rm H}=10^{21}$
cm$^{-2}$ and the dust-to-gas ratio is $f_d=f_\odot$.}
\label{fig:extnu}
\end{figure}

\begin{figure}
\centerline{\epsfysize=3.7in\epsffile{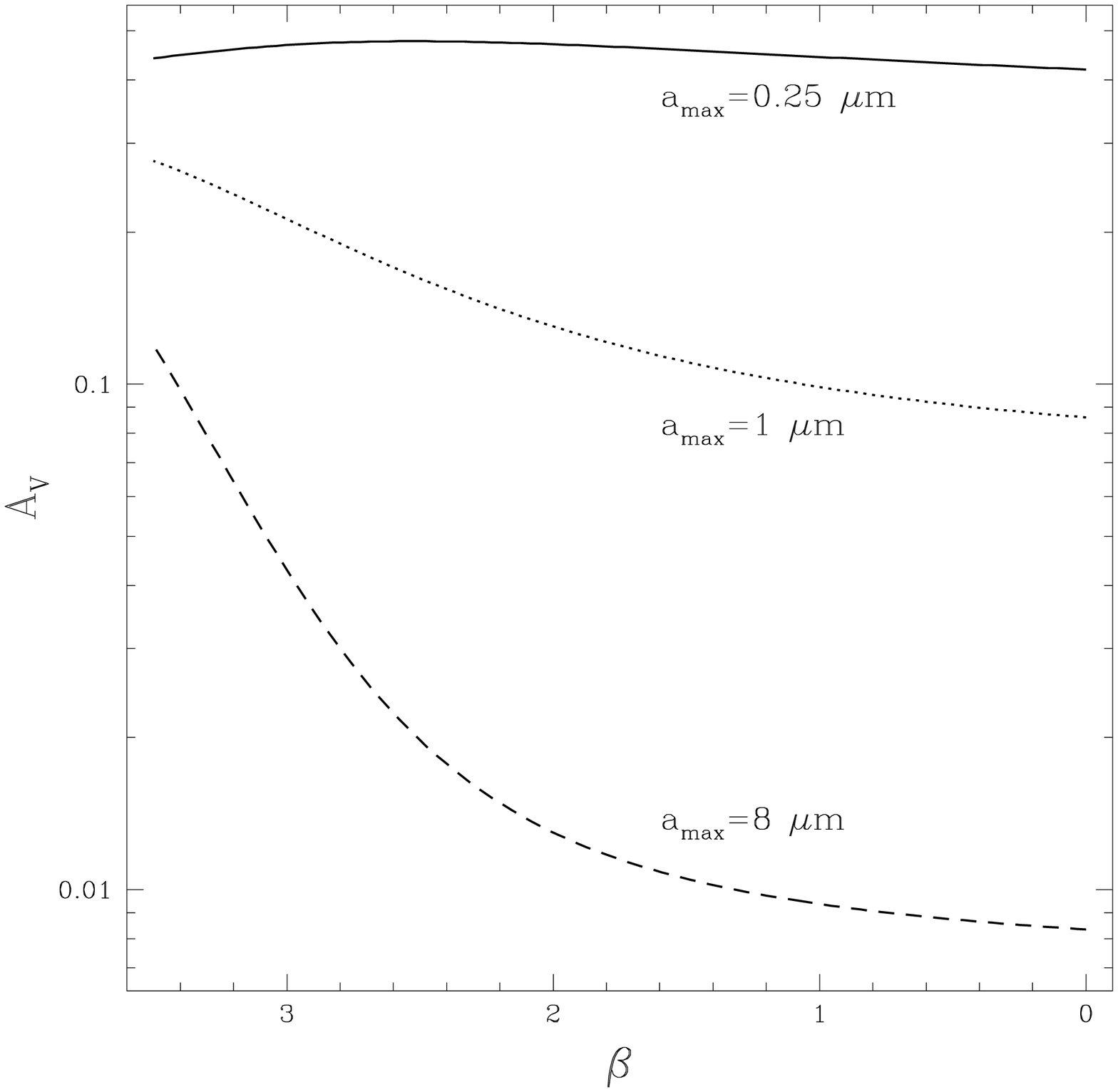}}
\centerline{\epsfysize=3.7in\epsffile{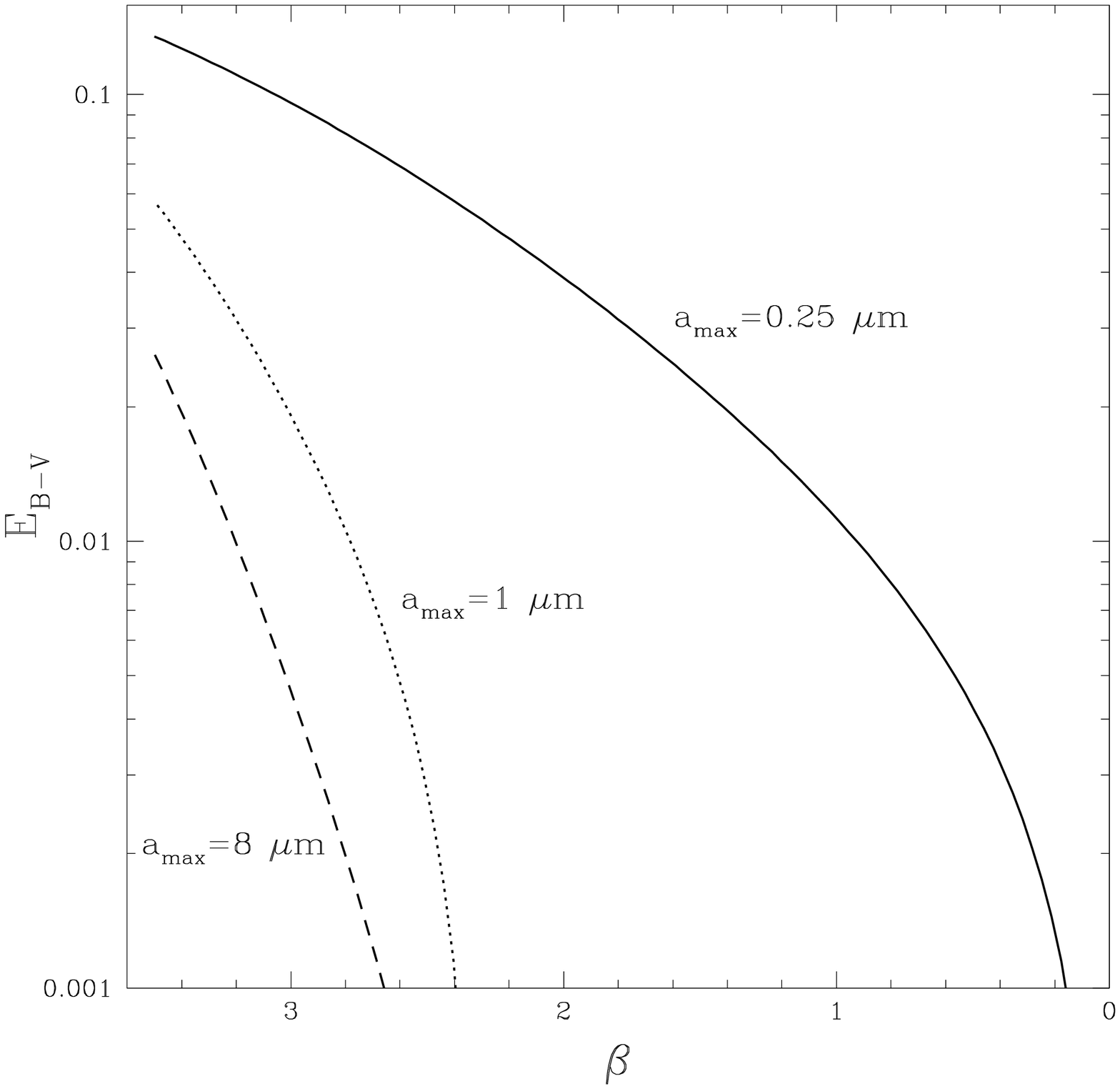}}
\caption{The V-band extinction (top panel) and the B-V reddening
(bottom panel) are shown as a function of the power-law slope
$\beta$ of the grain distribution. The grains are 
in the range \{$a_{\rm min}=0.005\, \mu{\rm m}$, $a_{\rm max}=0.25\,
\mu {\rm m}$\}. The Hydrogen column density is $N_{\rm H}=10^{21}$
cm$^{-2}$ and the dust-to-gas ratio is $f_d=f_\odot$. }
\label{fig:extbet}
\end{figure}

\begin{figure}
\centerline{\epsfysize=3.7in\epsffile{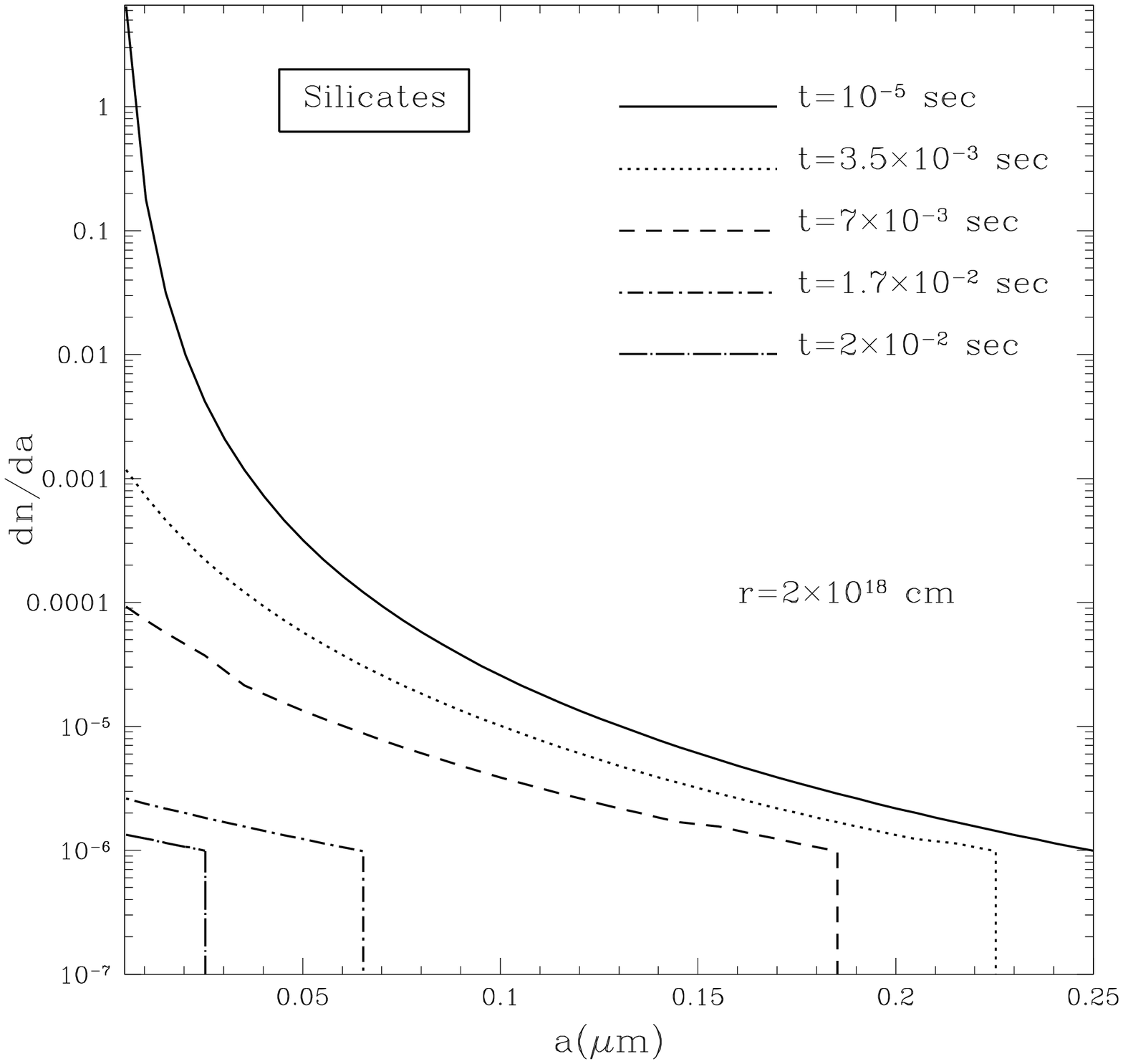}}
\centerline{\epsfysize=3.7in\epsffile{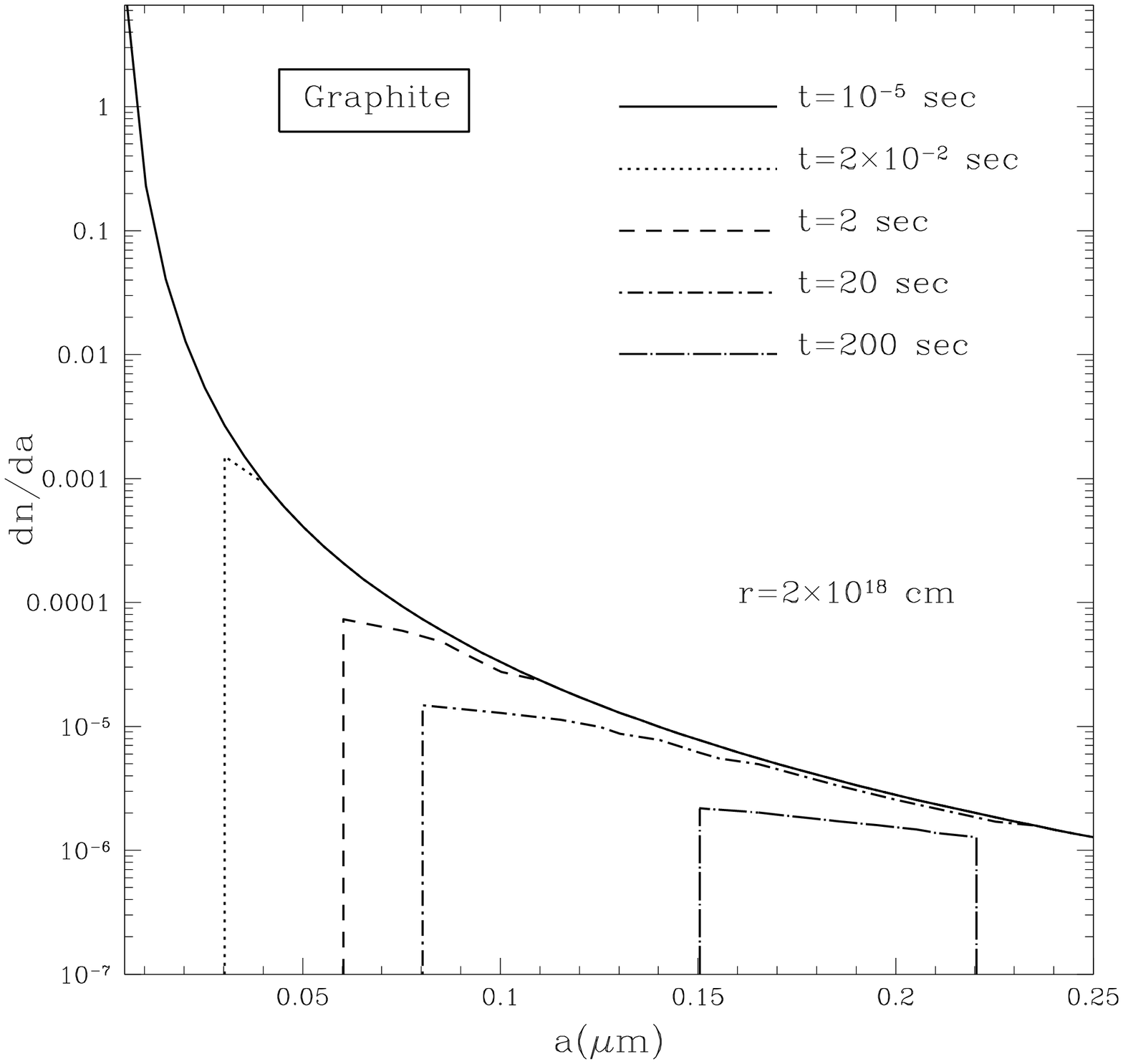}}
\caption{The evolution of the grain distribution at a distance
$r=2\times 10^{18}$ cm from the source as the radiation flux
impinges on the dust grains gradually reducing their sizes.
The top panel shows the evolution for the silicates, while the 
bottom panel for graphite. Note that the silicates evolve much
more rapidly than graphite.}
\label{fig:dndat}
\end{figure}

\begin{figure}
\centerline{\epsfysize=4in\epsffile{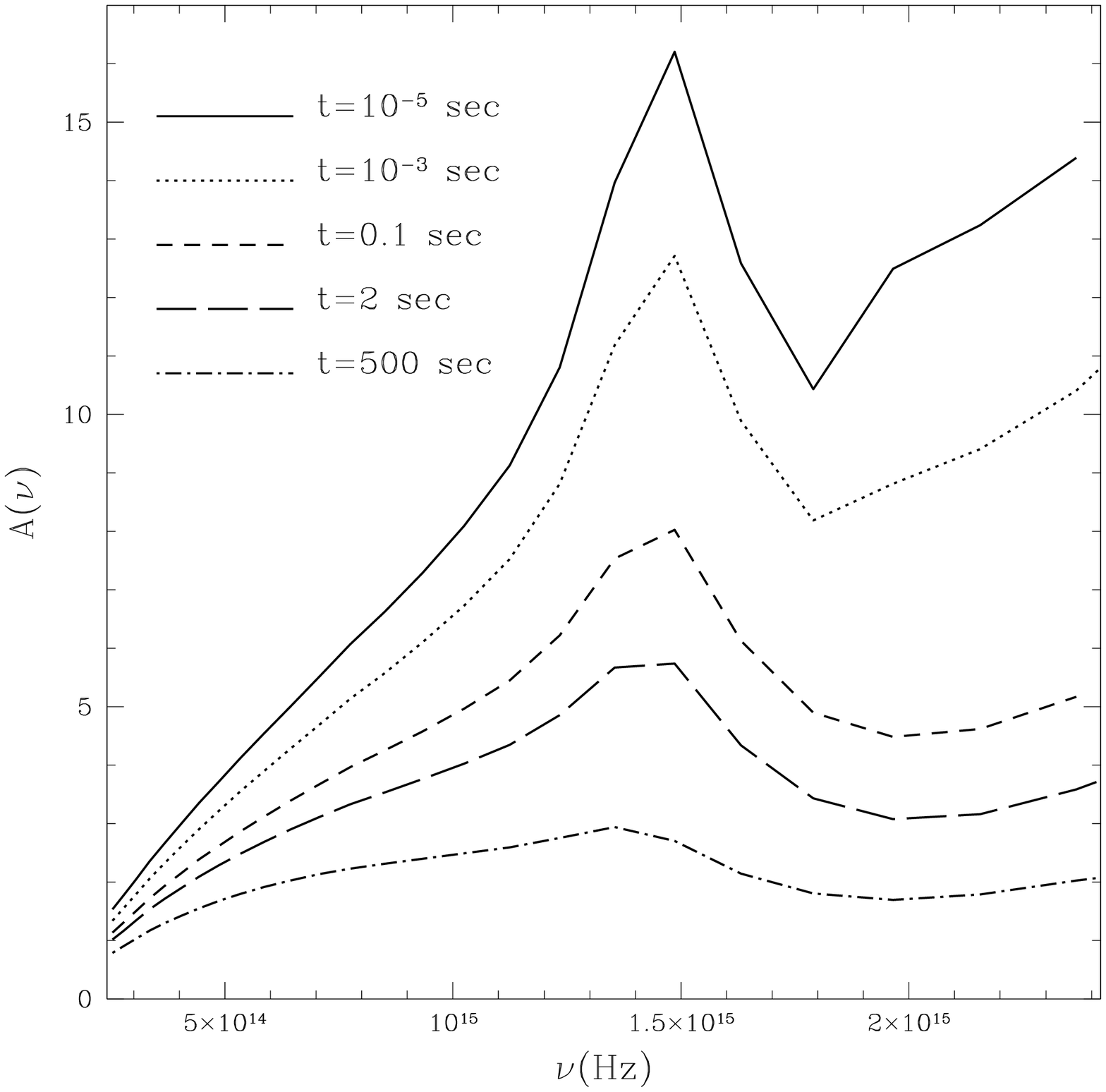}}
\centerline{\epsfysize=4in\epsffile{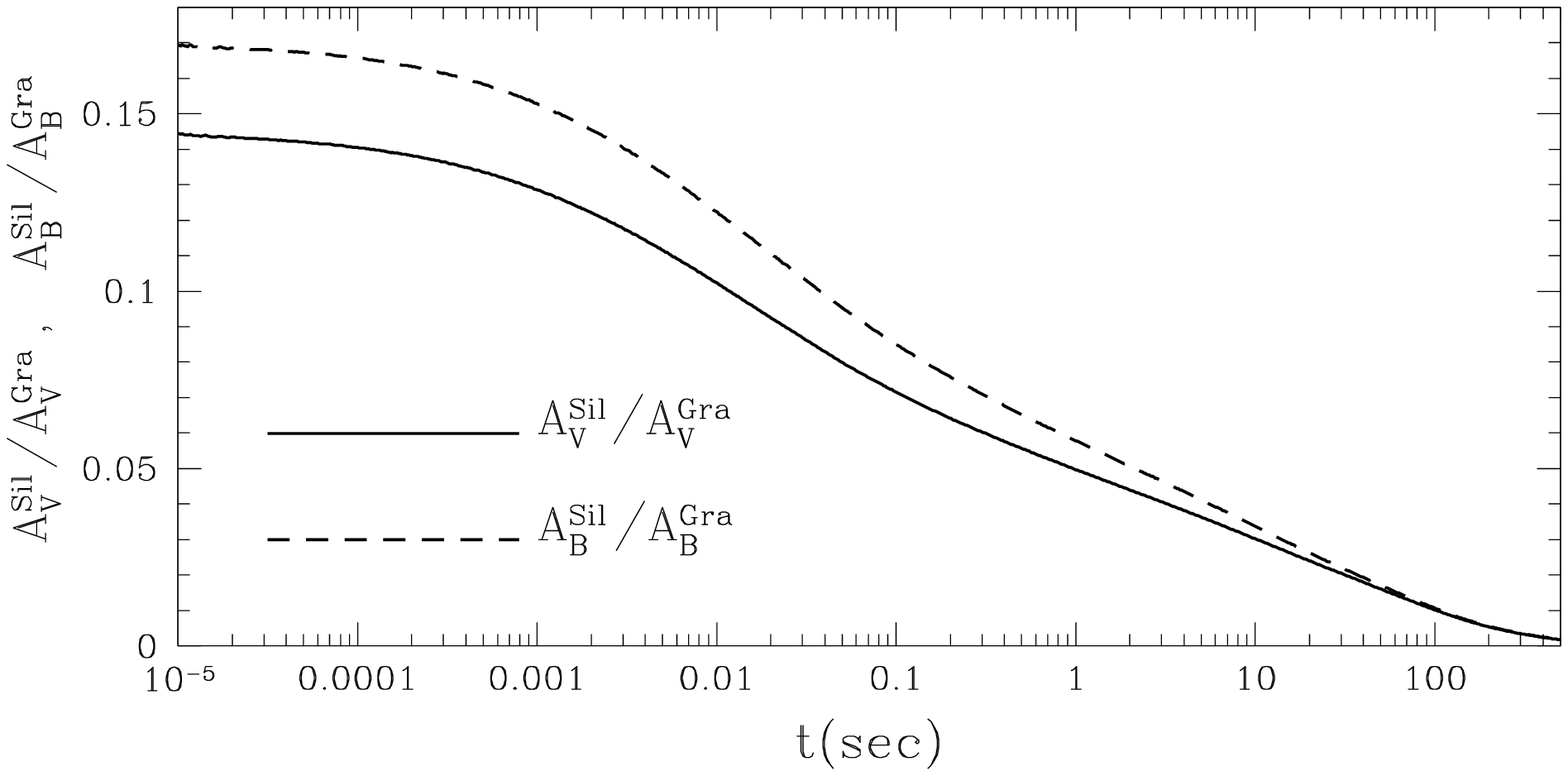}}
\caption{Top panel: the evolution of the extinction curve in a region
of size $R=10^{19}$ cm and density $n=10^3$ cm$^{-3}$, as the
radiation flux passes through it, altering the grain distribution.  At
$t=0$ the dust mass fraction in silicates and graphite is assumed to
be as for Galactic-type dust, but the bottom panel shows that the
contribution of graphite with respect to silicates increases
as the radiation flux impinges on the grains. }
\label{fig:extt}
\end{figure}

\begin{figure}
\centerline{\epsfysize=3.7in\epsffile{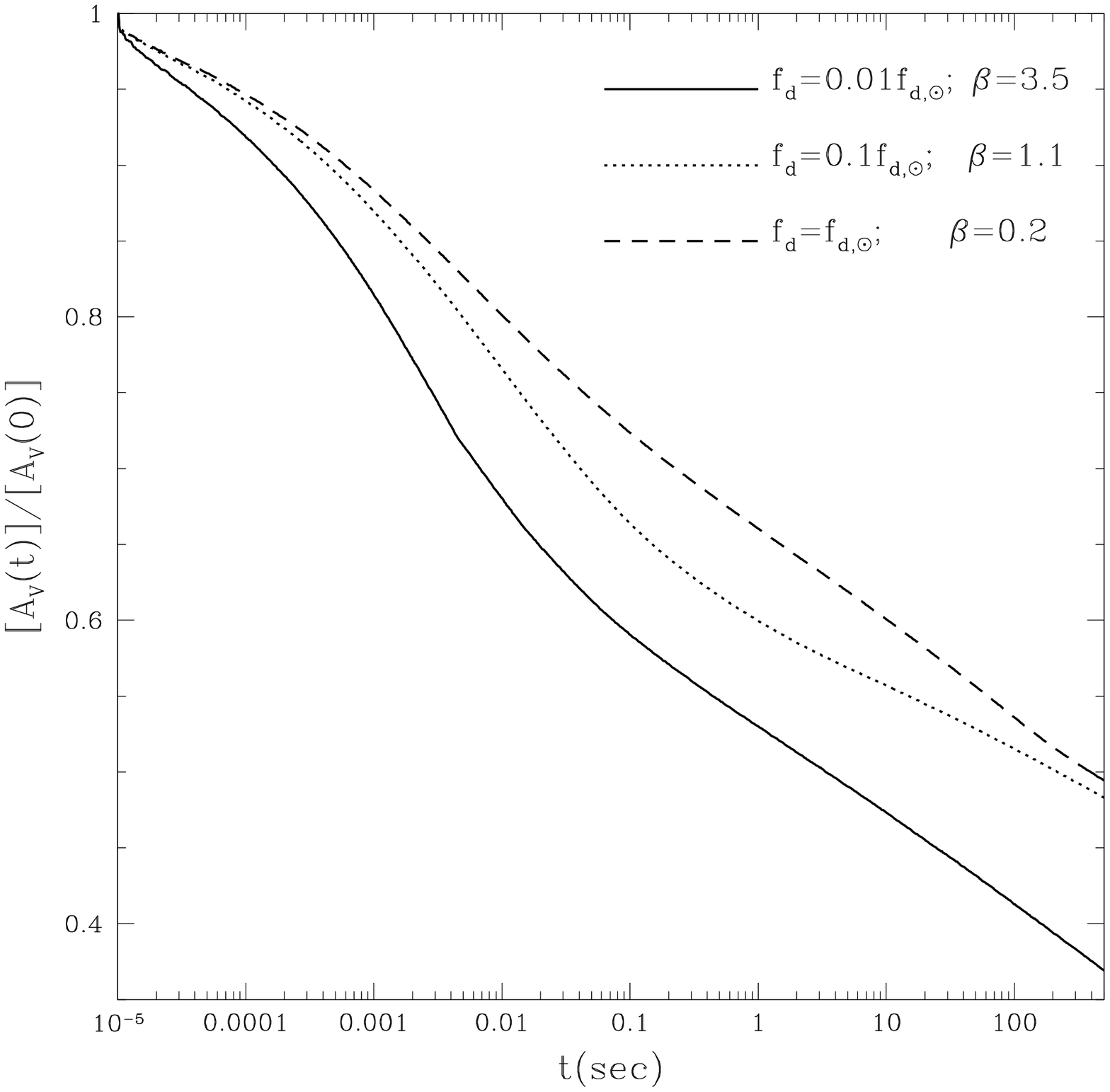}}
\centerline{\epsfysize=3.7in\epsffile{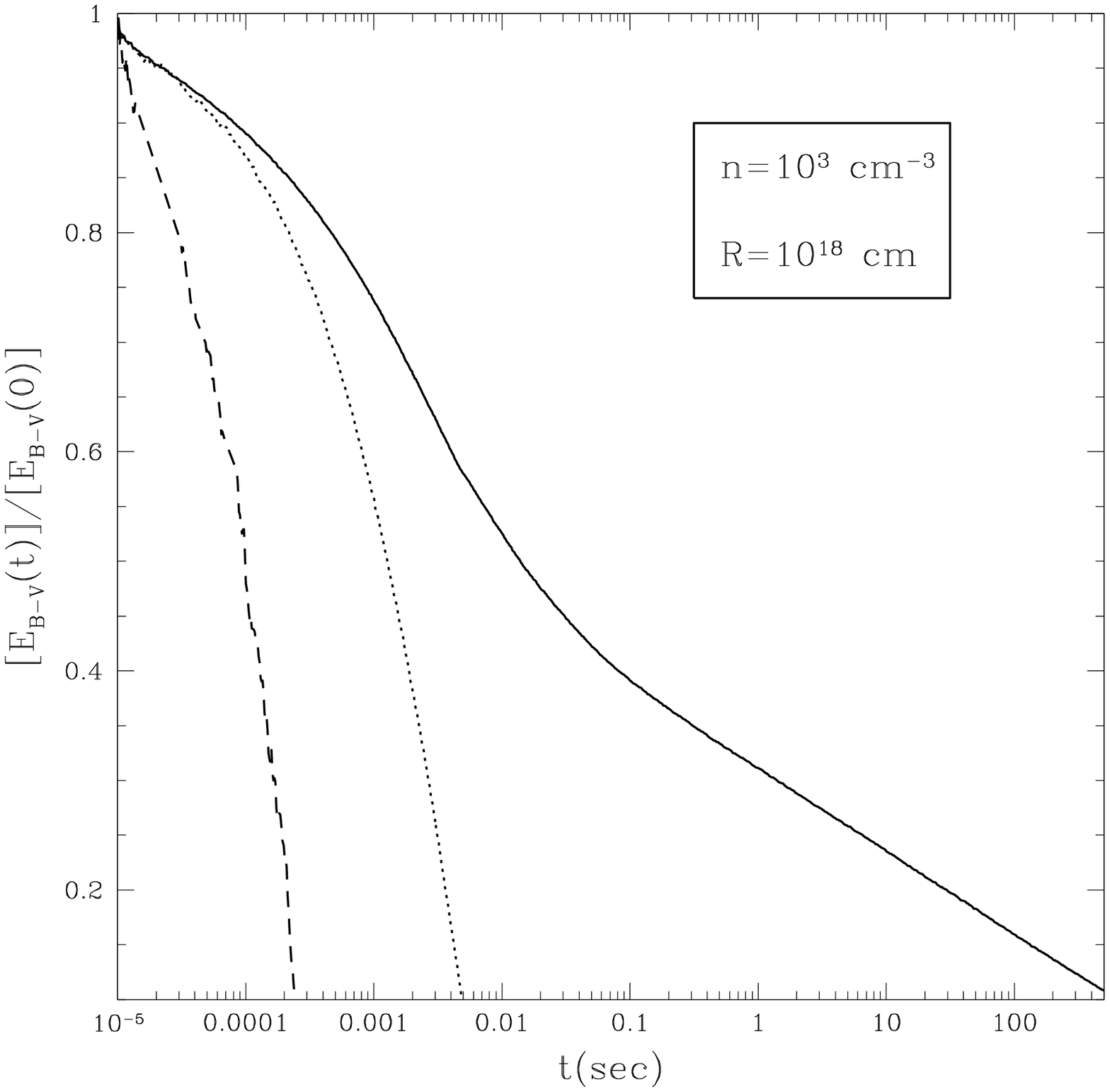}}
\caption{The behaviour of the extinction in the V band (top panel),
and reddening E$_{\rm B-V}$ (bottom panel) 
under an intense X-ray UV radiation field, is shown for
various slopes $\beta$ of the grain distribution and different values
of the dust-to-gas ratio $f_d$. The adopted values of \{$f_d$,
$\beta$\} are such that they yield the same reddening (before the 
source turns on). 
Note that the {\em steeper} the grain distribution, the
higher the degree of variability of the optical extinction, while
the {\em shallower} the grain distribution, the
higher the degree of variability of the reddening.}
\label{fig:avreddn1}
\end{figure}

\begin{figure}
\centerline{\epsfysize=3.7in\epsffile{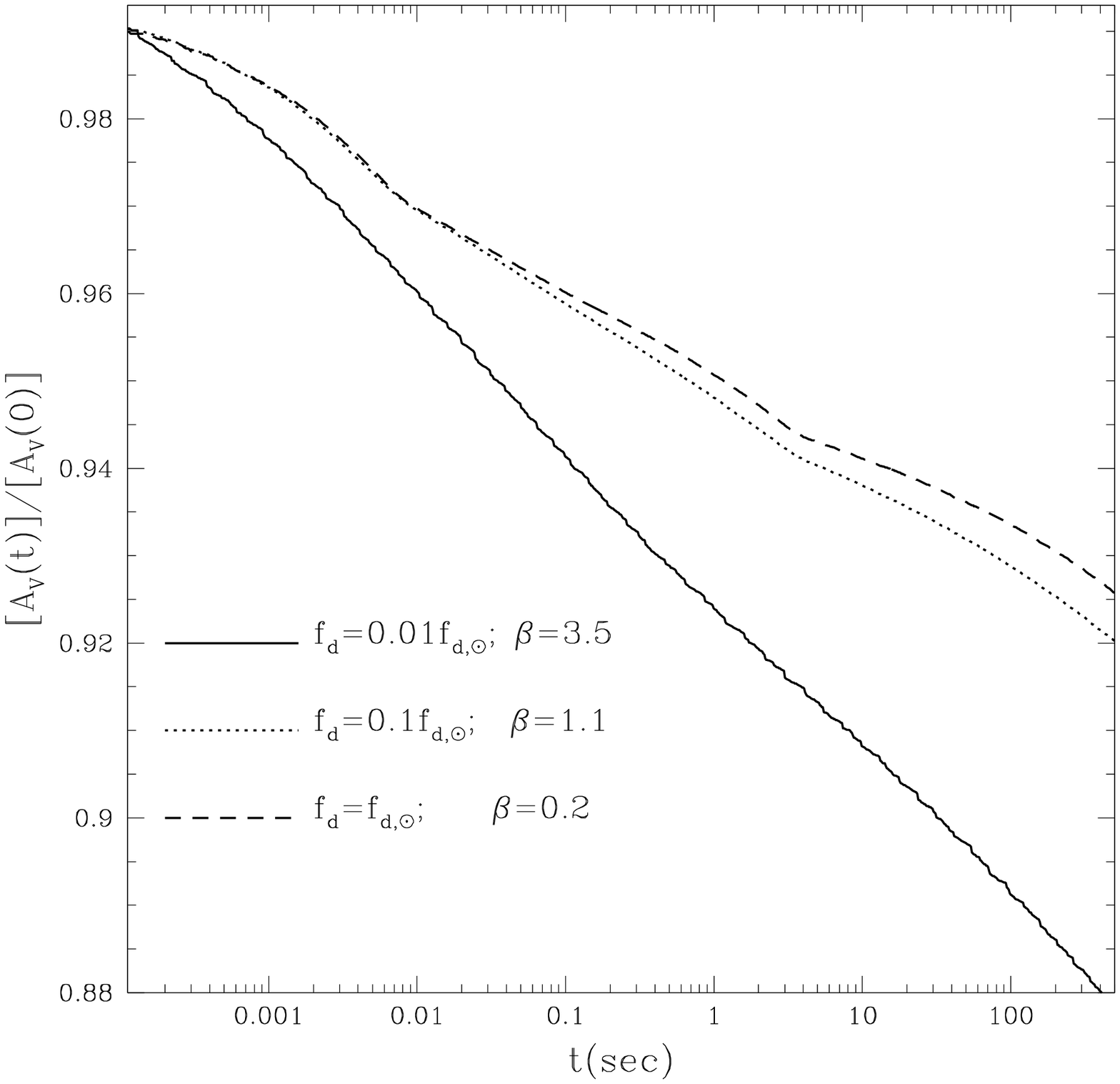}}
\centerline{\epsfysize=3.7in\epsffile{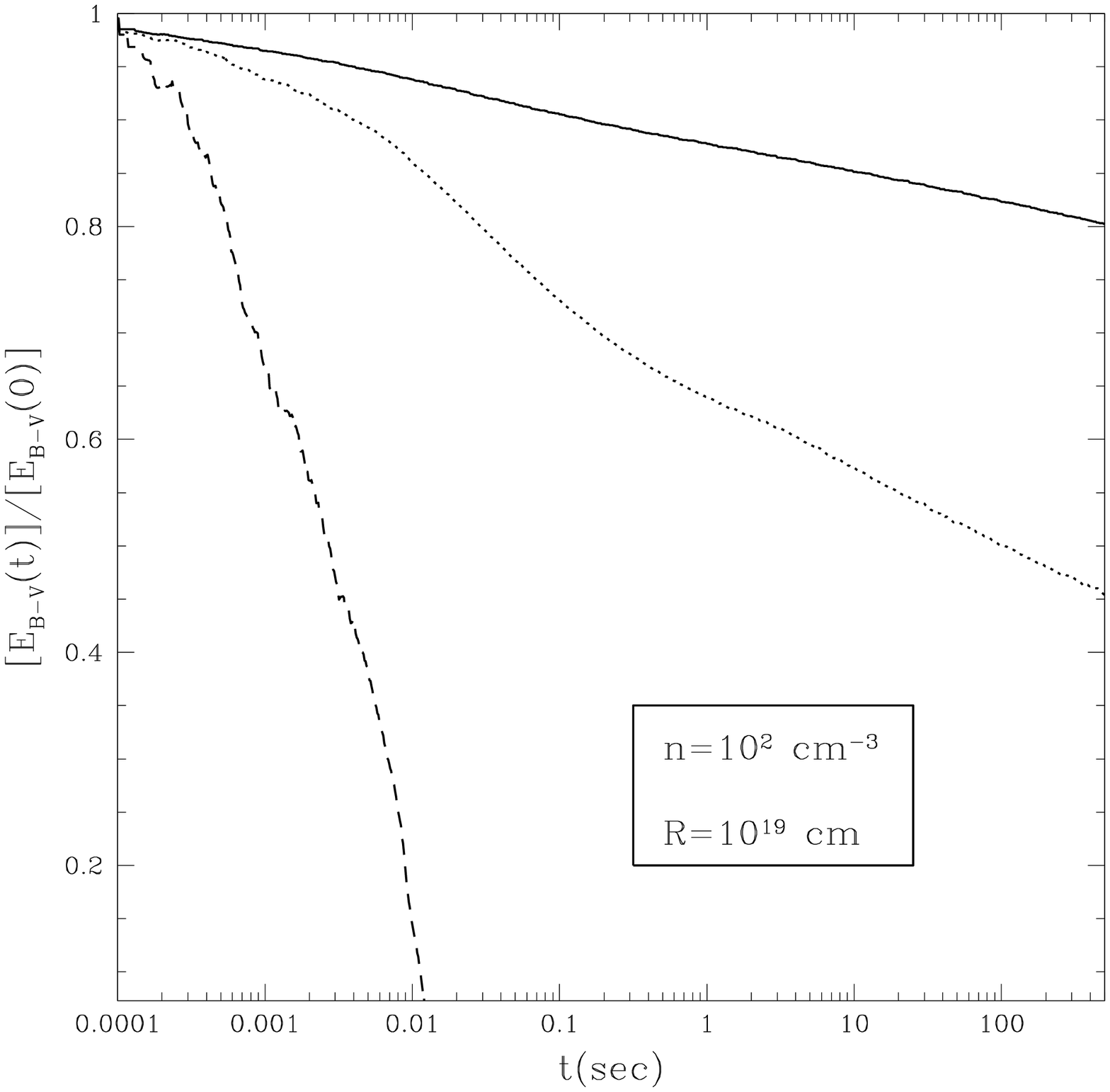}}
\caption{Same as in Fig. 5, but for a larger and less dense region.}
\label{fig:avreddn2}
\end{figure}

\begin{figure}
\centerline{\epsfysize=3.7in\epsffile{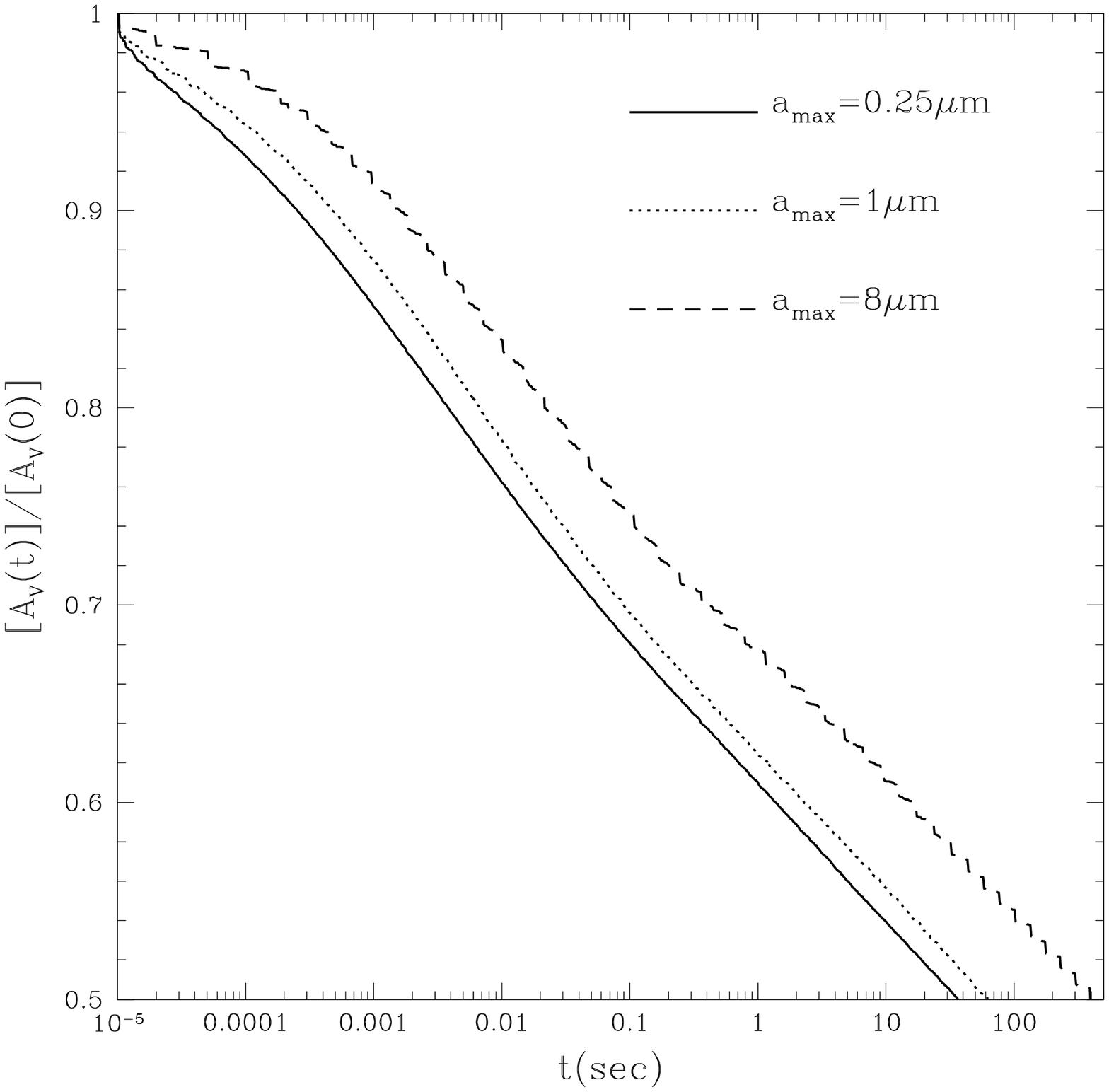}}
\centerline{\epsfysize=3.7in\epsffile{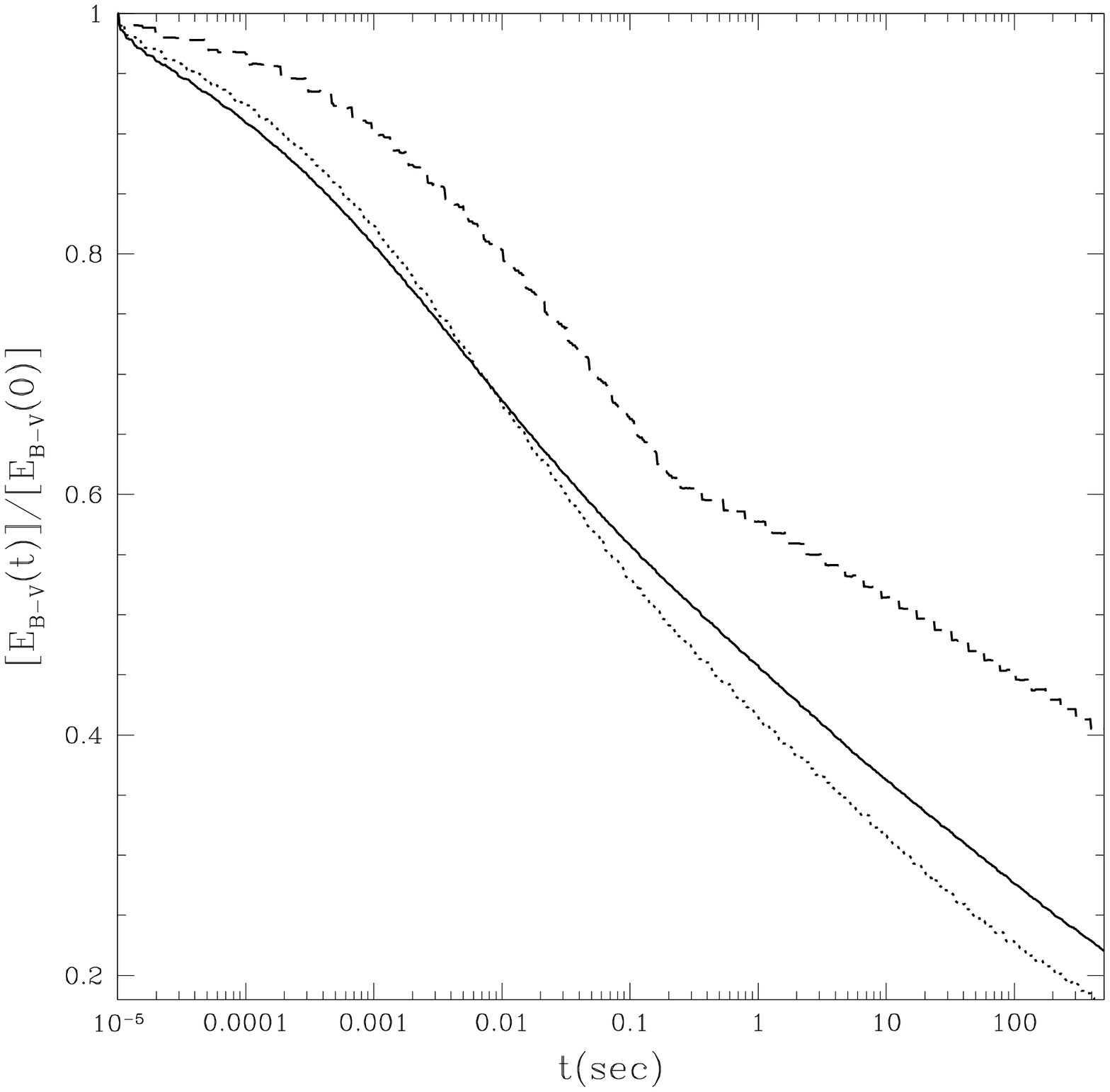}}
\caption{The behaviour of the extinction in the V band (top panel),
and reddening E$_{\rm B-V}$ (bottom panel) 
under the X-ray UV radiation field, is shown here for
various values of the largest grain $a_{\rm max}$ of the
initial dust grain distribution, and for the same 
value  $\beta=3.5$ of the slope of the distribution.}
\label{fig:avreddamax}
\end{figure}

\begin{figure}
\centerline{\epsfysize=5.7in\epsffile{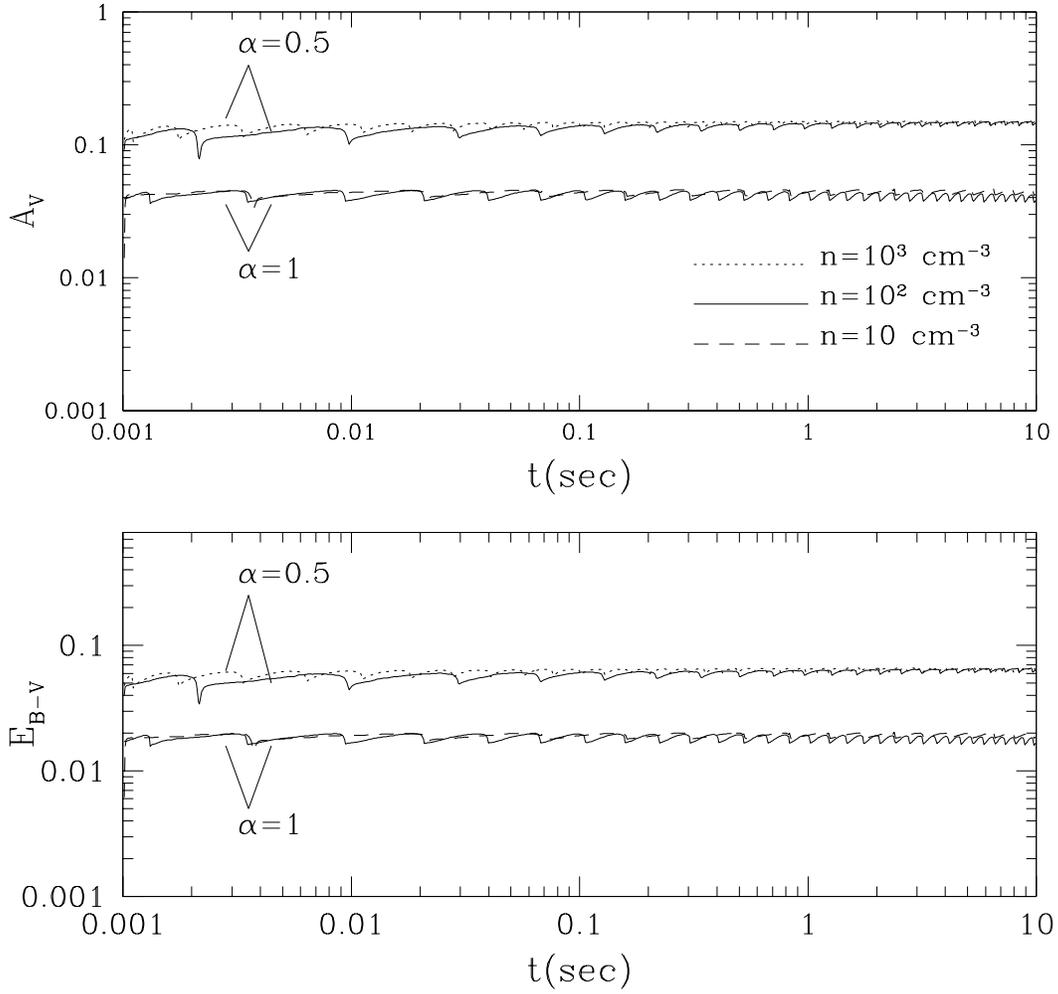}}
\caption{The contribution to the V-band opacity and the B-V reddening
due to the process of photodissociation of H$_2^+$. 
The oscillations are due to numerical inaccuracy in traversing radial shells.}
\label{fig:h2p}
\end{figure}

\begin{figure}
\centerline{\epsfysize=3.7in\epsffile{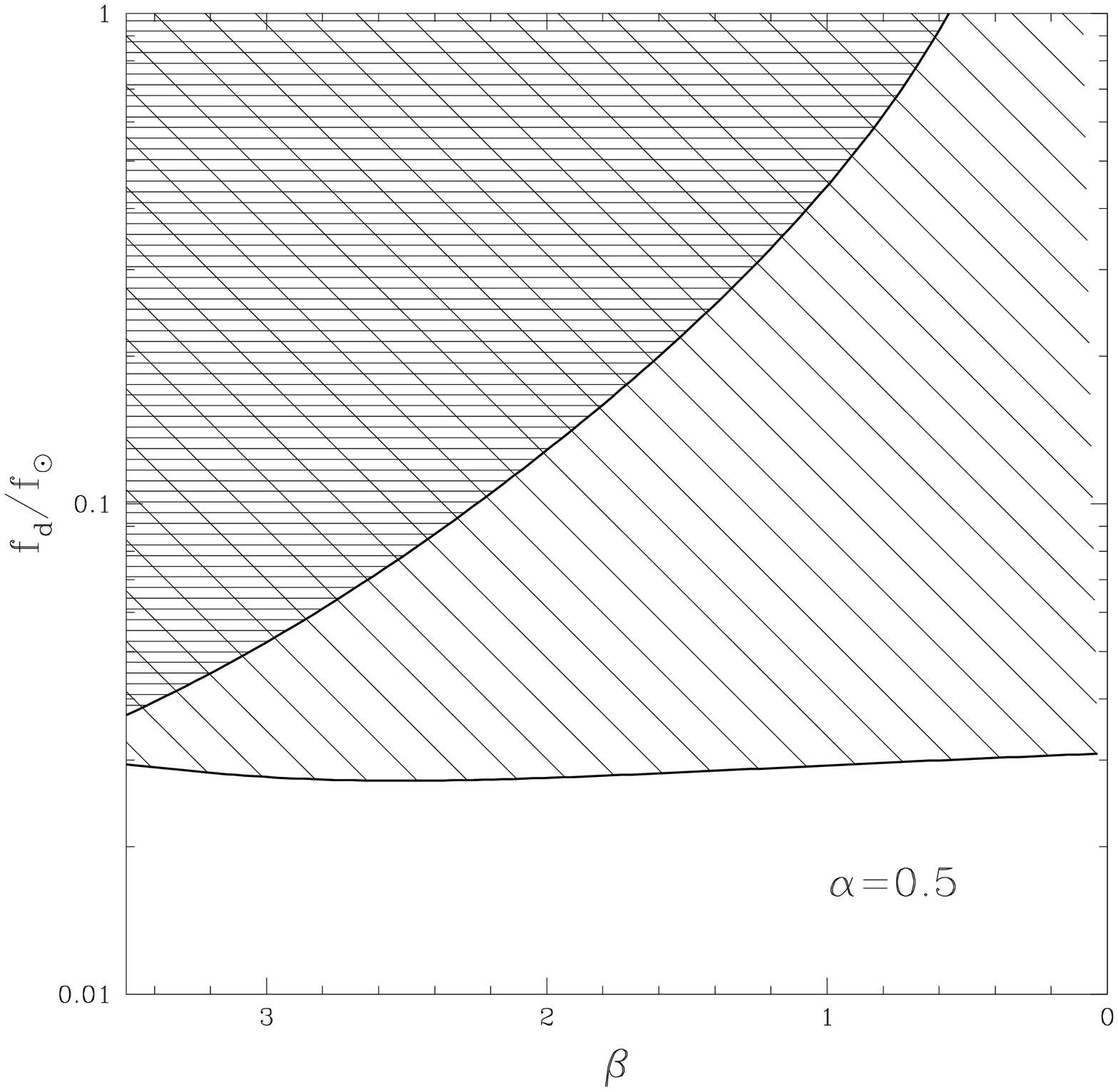}}
\centerline{\epsfysize=3.7in\epsffile{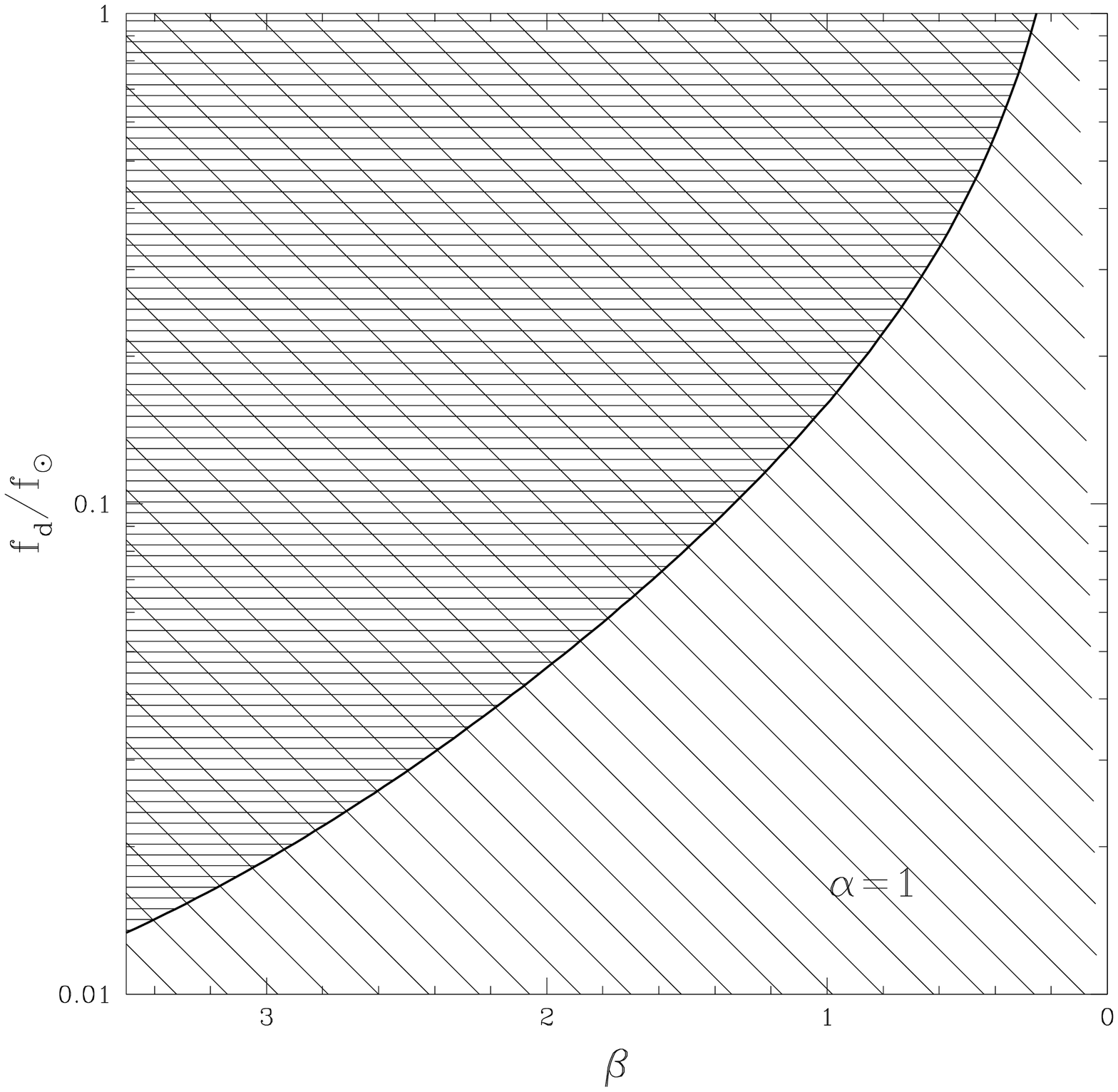}}
\caption{The shaded area marks the region in the parameter
space of \{$\beta\;,f_d$\} for which both the initial V-band opacity, $A^{\rm dust}_V(0)$, 
and reddening, E$^{\rm dust}_{\rm B-V}(0)$,  of dust 
(for a region whose Hydrogen 
column density is $N_{\rm H}=10^{22} {\rm cm}^{-2}$)
are higher than the contribution due to $H_2^+$. The dashed area in
between the two lines marks the region where $A^{\rm dust}_V(0)>A_V(H_2^+)$
but E$^{\rm dust}_{\rm B-V}(0)< {\rm E}_{\rm B-V}(H_2^+)$. }
\label{fig:h2lim}
\end{figure}

\begin{figure}
\centerline{\epsfysize=3.7in\epsffile{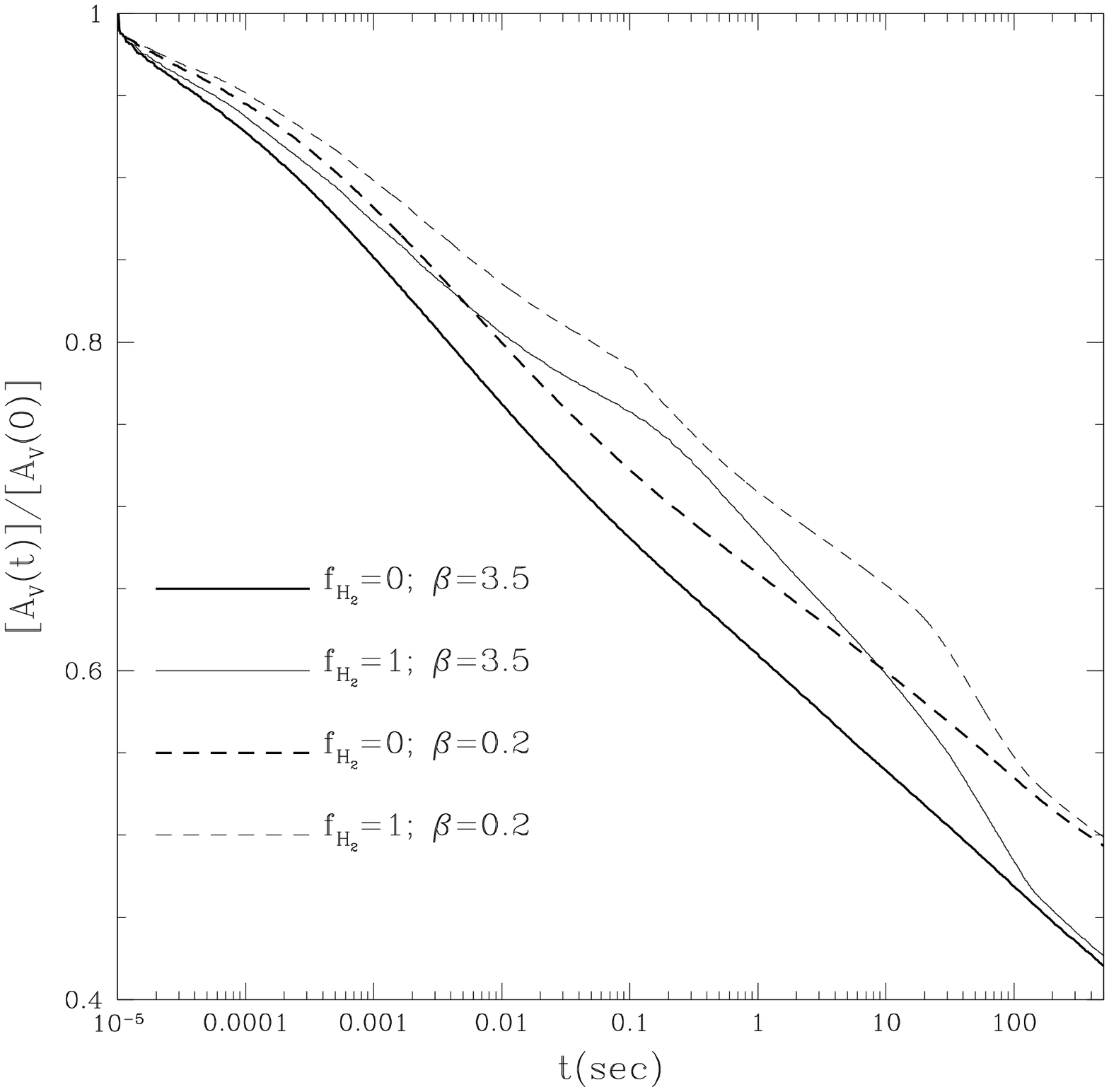}}
\centerline{\epsfysize=3.7in\epsffile{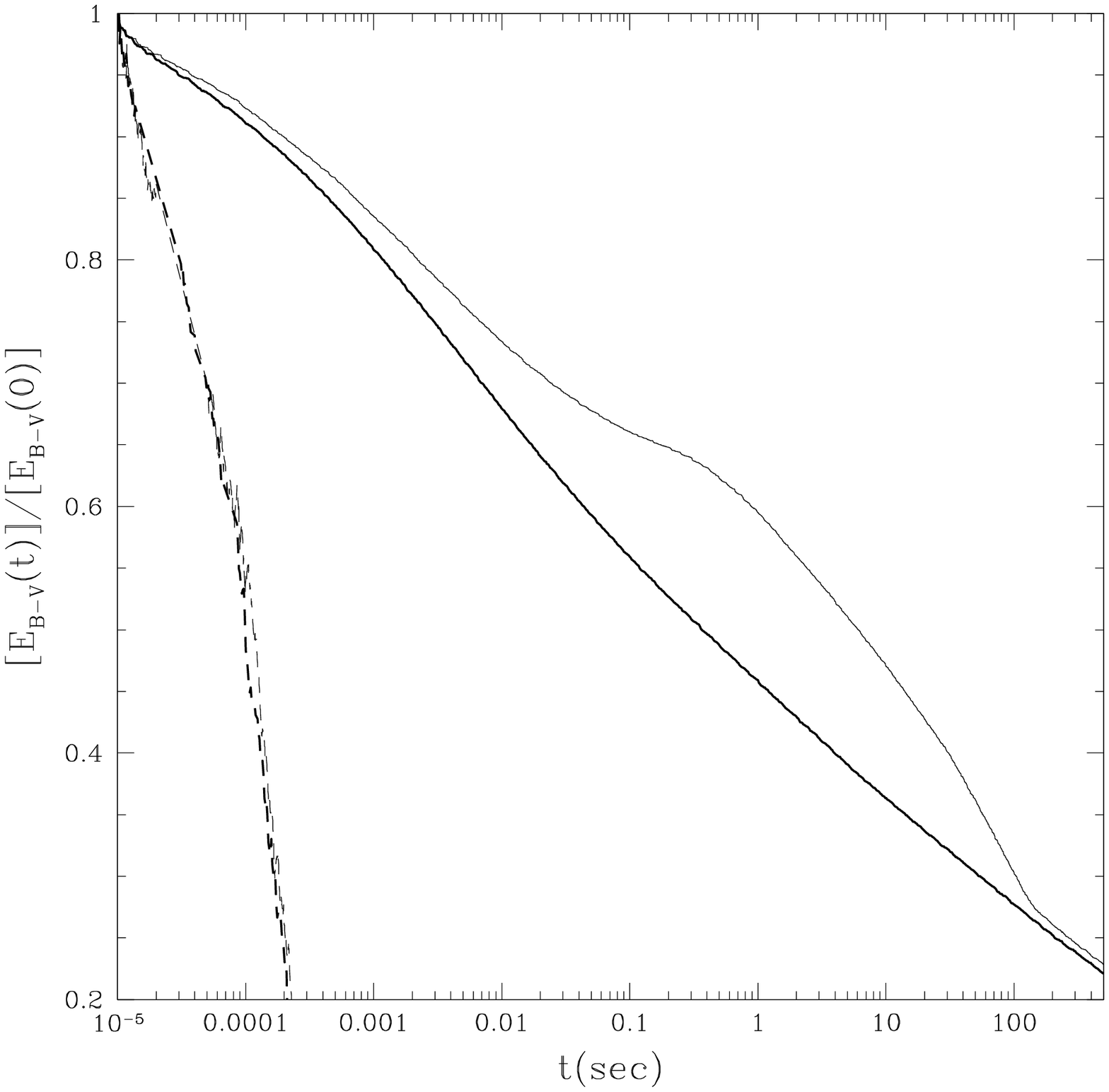}}
\caption{Comparison between the time evolution of the dust opacity and
reddening when all Hydrogen is initially in its atomic phase and when
it is all in its molecular phase. Here $f_{\rm H_2}$ represents the fraction of 
Hydrogen that is in molecular form.}
\label{fig:h2}
\end{figure}

\end{document}